\documentclass[12pt,preprint]{aastex}

\shorttitle{Two New Lensed Quasars from the SDSS}
\shortauthors{Inada et al.}

\begin{document}

\title{Two New Gravitationally Lensed Double Quasars from the
Sloan Digital Sky Survey} 

\author{
Naohisa Inada\altaffilmark{1,2},
Masamune Oguri\altaffilmark{3,4},
Robert H. Becker\altaffilmark{5,6},
Richard L. White\altaffilmark{7},
Issha Kayo\altaffilmark{8}, 
Christopher S. Kochanek\altaffilmark{9},
Patrick B. Hall\altaffilmark{10,4},
Donald P. Schneider\altaffilmark{11},
Donald G. York\altaffilmark{12,13}, 
and 
Gordon T. Richards\altaffilmark{14,15} 
}

\altaffiltext{1}{Institute of Astronomy, Faculty of Science,
  University of Tokyo, 2-21-1 Osawa, Mitaka, Tokyo 181-0015, Japan.}  
\altaffiltext{2}{Japan Society for the Promotion of Science (JSPS)
  Research Fellow.} 
\altaffiltext{3}{Kavli Institute for Particle Astrophysics and
  Cosmology, Stanford University, 2575 Sand Hill Road, Menlo Park, CA
  94025.} 
\altaffiltext{4}{Princeton University Observatory, Peyton Hall,
  Princeton, NJ 08544.} 
\altaffiltext{5}{IGPP-LLNL, L-413, 7000 East Avenue, Livermore, CA
  94550.} 
\altaffiltext{6}{Physics Department, University of California, Davis,
  CA 95616.} 
\altaffiltext{7}{Space Telescope Science Institute, 3700 San Martin
Drive, Baltimore, MD 21218.}  
\altaffiltext{8}{Department of Physics and Astrophysics, Nagoya
  University, Chikusa-ku, Nagoya 464-8062, Japan.}  
\altaffiltext{9}{Department of Astronomy, The Ohio State University, 
  Columbus, OH 43210.}
\altaffiltext{10}{Department of Physics and Astronomy, York
  University, 4700 Keele Street, Toronto, Ontario, M3J 1P3, Canada.}
\altaffiltext{11}{Department of Astronomy and Astrophysics, The
  Pennsylvania State University, 525 Davey Laboratory, University
  Park, PA 16802.}   
\altaffiltext{12}{Department of Astronomy and Astrophysics, The
  University of Chicago,5640 South Ellis Avenue, Chicago, IL 60637.} 
\altaffiltext{13}{Enrico Fermi Institute, The University of Chicago,
  5640 South Ellis Avenue, Chicago, IL 60637.}
\altaffiltext{14}{Johns Hopkins University, 3400 N. Charles St.,
  Baltimore, MD 21218.} 
\altaffiltext{15}{Department of Physics, Drexel University, 3141
  Chestnut Street, Philadelphia, PA 19104.}

\begin{abstract}
We report the discoveries of the two-image gravitationally lensed
quasars, SDSS J0746+4403 and SDSS J1406+6126, selected from the Sloan
Digital Sky Survey (SDSS). SDSS J0746+4403, which will be included in
our lens sample for statistics and cosmology, 
has a source redshift of $z_s=2.00$, an estimated lens redshift of 
$z_l\sim 0.3$, and an image separation of $1\farcs08$. 
SDSS J1406+6126 has a source redshift of $z_s=2.13$, 
a spectroscopically measured lens redshift of $z_l=0.27$, and an 
image separation of $1\farcs98$. We find that the two quasar images of
SDSS J1406+6126 have different intervening \ion{Mg}{2} absorption
strengths, which are suggestive of large variations of absorbers on 
kpc scales. The positions and fluxes of both the lensed quasar systems 
are easily reproduced by simple mass models with reasonable parameter 
values. These objects bring to 18 the number of lensed quasars that have 
been discovered from the SDSS data.
\end{abstract}


%%%%%%%%%%%%%%%%%%%%%%%%%%%%%%%%%%%%%%%%%%%%%%%%
%%%%%%%%%%%%%%%%%%%%%%%%%%%%%%%%%%%%%%%%%%%%%%%%
%%%%%%%%%%%%%%%%%%%%%%%%%%%%%%%%%%%%%%%%%%%%%%%%
\section{Introduction}\label{sec:intro}
%%%%%%%%%%%%%%%%%%%%%%%%%%%%%%%%%%%%%%%%%%%%%%%%
%%%%%%%%%%%%%%%%%%%%%%%%%%%%%%%%%%%%%%%%%%%%%%%%
%%%%%%%%%%%%%%%%%%%%%%%%%%%%%%%%%%%%%%%%%%%%%%%%

Gravitationally lensed quasars are unique astronomical
and cosmological tools \citep[see the review by][]{kochanek06}. 
They are useful for studying the mass distributions, substructures and 
interstellar mediums of their lens galaxies, the structure of the
quasar host galaxy and accretion disk, and estimating the cosmological
model \citep[e.g.,][]{kochanek91,mao98,chang84,turner90,fukugita90,
refsdal64}. Lensed quasars due to foreground galaxy clusters 
probe the distribution of dark matter in the lens objects 
\citep{oguri04a,williams04}, and their statistics can be an useful 
test of dark matter models \citep{narayan88,oguri04b}. A continuing
problem, however, is the limited number of strongly lensed quasars
\citep[approximately 100,][]{kochanek06} and the still more limited
number found in systematic surveys. In particular, the current largest
completed survey, the Cosmic Lens All Sky Survey
\citep[CLASS;][]{myers03,browne03}, led to a well-defined statistical
sample of only 13 lensed radio sources. While the CLASS survey can be
used to obtain constraints on dark energy \citep{chae02,mitchel2005},
larger well-defined lens samples are needed to be competitive with
other approaches to constraining the cosmological model. 

The Sloan Digital Sky Survey \citep[SDSS;][]{york00} is currently the 
best candidate for building a larger, well-defined lensed quasar sample 
than the CLASS. The SDSS is expected to eventually provide a catalog of 
roughly $100,000$ spectroscopically identified quasars
\citep{schneider05,schneider06}, which we have been searching for
gravitationally lensed quasars in the SDSS Quasar Lens Search
\citep[SQLS;][]{oguri06,inada06b}. Since the probability of a quasar
being multiply imaged is of order $10^{-3}$ \citep{turner84}, we have
a reasonable expectation of obtaining a well-defined sample of roughly
100 lensed quasars. Including the two new lenses reported here, we
have now discovered 18 new lensed quasars \citep{inada03a,inada03b,
inada03c,inada05,inada06a,johnston03,morgan03,pindor04,pindor05,
oguri04c,oguri05,burles06,morokuma06,kayo06}, and recovered 8
previously known lensed quasars \citep{walsh79,weymann80,surdej87,
bade97,oscoz97,schechter98,morgan01,magain88}, making the SQLS the
largest survey for gravitational lensed quasars to date.  

In this paper, we present two new doubly-imaged lensed quasars found 
in the course of the SQLS, SDSS J074653.03+440351.3 (SDSS J0746+4403) 
and SDSS J140624.82+612640.9 (SDSS J1406+6126). They were confirmed 
with the observations at the W. M. Keck Observatory's Keck II
telescope, the University of Hawaii 2.2-meter (UH88) telescope, and the
MDM Observatory's 2.4-m Hiltner (MDM 2.4-m) telescope. The structure of
this paper is as follows. We describe our lens candidate selection from
the SDSS data in \S \ref{sec:sdss}. The results of the imaging and
spectroscopic follow-up observations for SDSS J0746+4403 and SDSS
J1406+6126 are presented in \S \ref{sec:0746} and \S \ref{sec:1406},
respectively.  Section \ref{sec:model} discusses the models of the
lensed quasars and we summarize our results in \S \ref{sec:conc}. 
We assume a cosmological model with matter density $\Omega_M=0.27$, 
cosmological constant $\Omega_\Lambda=0.73$, and Hubble constant 
$h=H_0/100{\rm km\,sec^{-1}Mpc^{-1}}=0.7$ \citep{spergel03}. 

%%%%%%%%%%%%%%%%%%%%%%%%%%%%%%%%%%%%%%%%%%%%%%%%
%%%%%%%%%%%%%%%%%%%%%%%%%%%%%%%%%%%%%%%%%%%%%%%%
%%%%%%%%%%%%%%%%%%%%%%%%%%%%%%%%%%%%%%%%%%%%%%%%
\section{Selection Algorithm}\label{sec:sdss}
%%%%%%%%%%%%%%%%%%%%%%%%%%%%%%%%%%%%%%%%%%%%%%%%
%%%%%%%%%%%%%%%%%%%%%%%%%%%%%%%%%%%%%%%%%%%%%%%%
%%%%%%%%%%%%%%%%%%%%%%%%%%%%%%%%%%%%%%%%%%%%%%%%

The SDSS consists of a photometric survey
\citep{gunn98,lupton99,tucker06} in five broad-band optical filters
\citep{fukugita96} and a spectroscopic survey with a multi-fiber
spectrograph covering 3800{\,\AA} to 9200{\,\AA} at a resolution of
$\hbox{R}\sim1800$, both using a dedicated wide-field 
($3^{\circ}$ field of view)
2.5-m telescope \citep{gunn06} at the Apache Point Observatory in New
Mexico, USA. The final survey area is about 10,000 square degrees of
the sky approximately centered on the North Galactic Cap. Once the
imaging data is processed by the photometric pipeline
\citep{lupton01,lupton05}, quasar and galaxy candidates are selected
for spectroscopy \citep{eisenstein01,richards02,strauss02} and the
spectroscopic observations are conducted according to the tiling
algorithm of \citet{blanton03}. The imaging data has an astrometric
accuracy better than about $0\farcs1$ rms per coordinate
\citep{pier03} and photometric zeropoint errors less than about 0.03
magnitude over the entire survey area
\citep{hogg01,smith02,ivezic04}. The SDSS is continuously releasing
the data to the public 
\citep{stoughton02,abazajian03,abazajian04,abazajian05,adelman06}.   

The two objects, SDSS J0746+4403 and SDSS J1406+6126, were selected as  
lensed quasar candidates from the sample of the SDSS spectroscopically 
confirmed quasars by two algorithms. The first method is the algorithm  
we used previously \citep[e.g.,][]{inada03a}, which identifies 
close separation ($\sim$1\farcs0) lens candidates as quasars with low
probabilities of being well fit as a point source, an exponential disk
or a de Vaucouleurs profile. The second is the new algorithm
introduced by \citet{oguri06} that examines close pairs of objects
(quasars); both candidates also satisfy the new morphological
selection criteria.  We note that SDSS J1406+6126 with $i_{\rm cor}=19.31$ 
($i$-band Point-Spread Function magnitude corrected for Galactic
extinction) is slightly fainter than the magnitude limit we are using
to construct our statistical lens sample \citep[see][]{oguri06}.  

\begin{figure}
\epsscale{.45}
\plotone{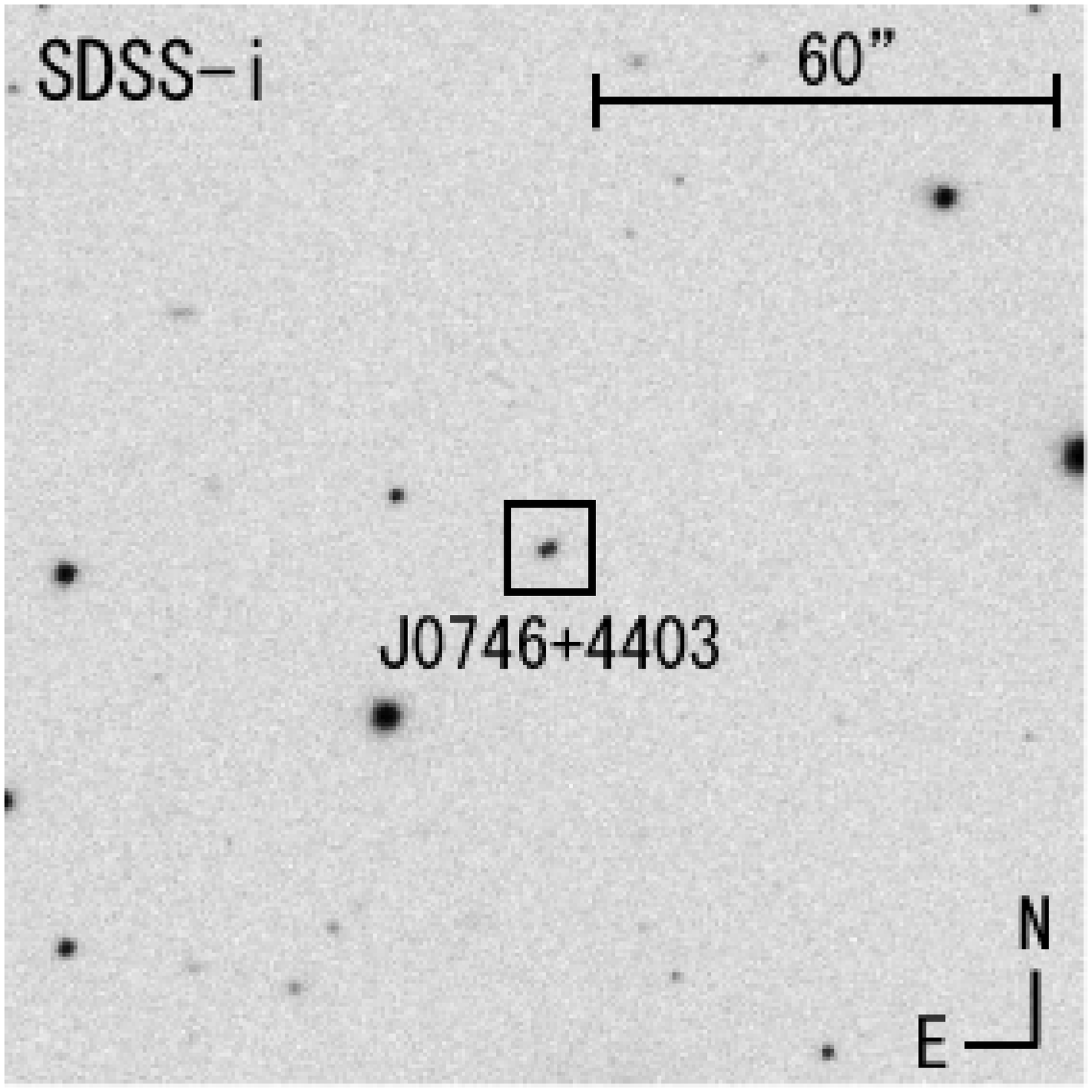}
\plotone{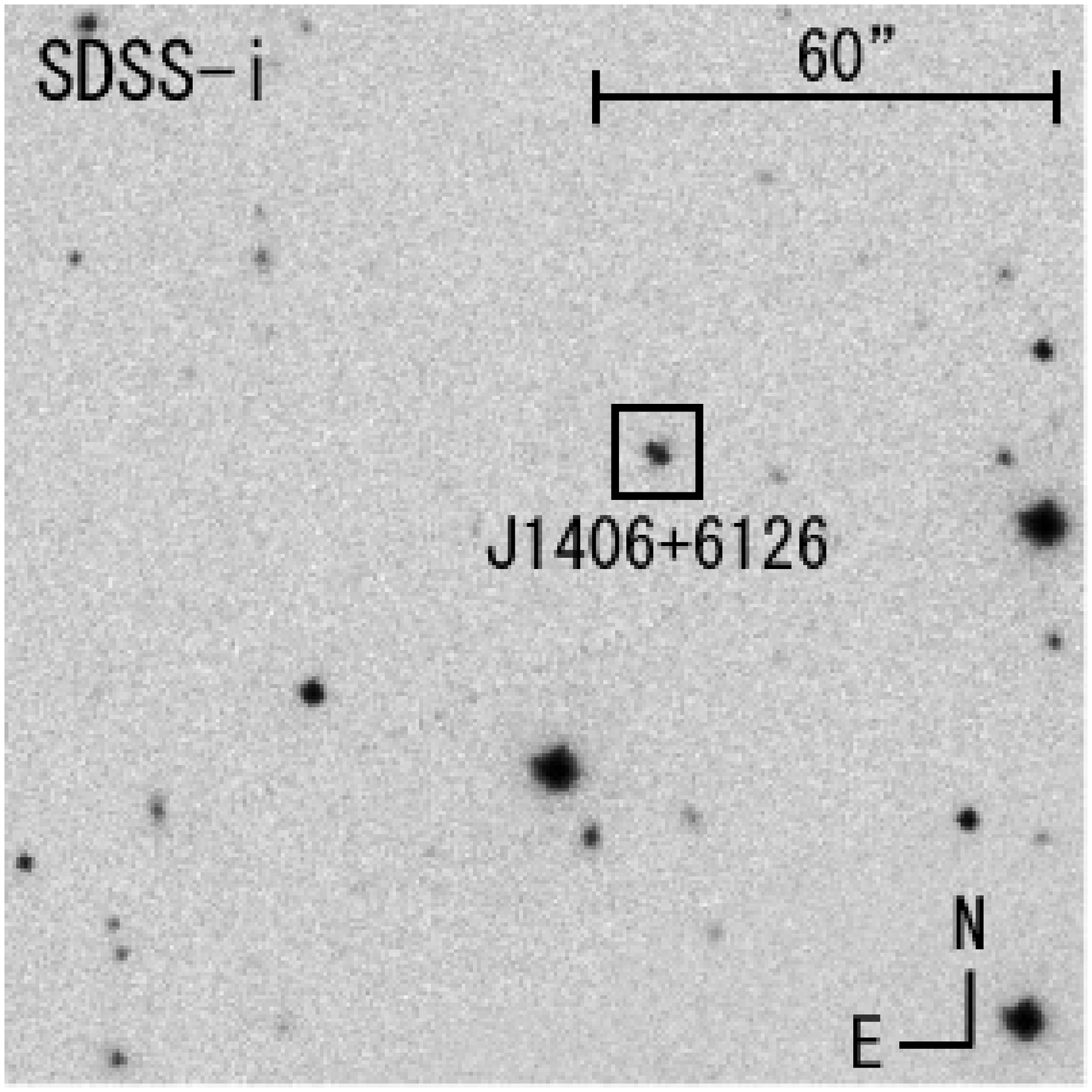}
\caption{SDSS $i$-band finding charts for SDSS J0746+4403
({\it left}) and SDSS J1406+6126 ({\it right}). The image scale
 is 0\farcs396 ${\rm pixel^{-1}}$, North is up and East is left.
\label{fig:findingchart}}
\end{figure}

Figure~\ref{fig:findingchart} shows the SDSS $i$-band images of
SDSS J0746+4403 and SDSS J1406+6126. The total magnitudes 
inside a ${\sim}2\farcs0$ aperture radius (without Galactic 
extinction corrections) and the quasar redshifts from the 
SDSS imaging and spectroscopic data are summarized in 
Table~\ref{tab:sdss}. Both objects are marginally resolved in the 
SDSS imaging data and  (spatially) unresolved in the SDSS
spectroscopic data, thus additional observations were needed to
confirm that they are indeed gravitationally lensed quasars. 

%%%%%%%%%%%%%%%%%%%%%%%%%%%%%%%%%%%%%%%%%%%%%%%%
%%%%%%%%%%%%%%%%%%%%%%%%%%%%%%%%%%%%%%%%%%%%%%%%
%%%%%%%%%%%%%%%%%%%%%%%%%%%%%%%%%%%%%%%%%%%%%%%%
\section{Follow-Up Observations of  SDSS~J0746+4403}\label{sec:0746}
%%%%%%%%%%%%%%%%%%%%%%%%%%%%%%%%%%%%%%%%%%%%%%%%
%%%%%%%%%%%%%%%%%%%%%%%%%%%%%%%%%%%%%%%%%%%%%%%%
%%%%%%%%%%%%%%%%%%%%%%%%%%%%%%%%%%%%%%%%%%%%%%%%

We obtained $V$-, $R$-, and $I$-band images of SDSS J0746+4403 with the 
8k mosaic CCD camera (UH8k, 0\farcs235 pixels) and the Orthogonal 
Parallel Transfer Imaging Camera (OPTIC, 0\farcs137 pixels) at the 
UH88 telescope.  We also obtained a deep $r$-band image with the 
Retractable Optical Camera for Monitoring \citep[RETROCAM, 0\farcs259
pixels,][]{morgan05} at the MDM 2.4-m telescope.  The observations were
conducted on 2004 December 14 (UH8k-$V$, 0\farcs7 seeing, 240 sec), 
2004 December 16 (UH8k-$R$ and -$I$, 0\farcs8 seeing, 360 and 270 sec), 
2006 February 23 (RETROCAM-$r$, 1\farcs0 seeing, 2700 sec), and 2006 May
3  (OPTIC-$I$, 1\farcs0 seeing, 300 sec). In all these images, 
which are shown in the left column of Figure~\ref{fig:0746img}, we
observe two stellar components (denoted as A and B; A being the 
Eastern component) with a separation of $1\farcs079\pm0\farcs010$. The
middle column of Figure~\ref{fig:0746img} shows the residuals after
fitting the images using {\it only} two point spread functions (PSFs)
constructed from nearby stars. In each case, there are significant,
extended residual fluxes (named G),  particularly in the redder
bands. The most natural interpretation is that  they come from a lens
galaxy. The right column of Figure~\ref{fig:0746img} shows the
residuals after using GALFIT \citep{peng02} to fit the data with two
PSFs and an extended galaxy (modeled by a S\'{e}rsic profile)  
between them.  With the addition of a galaxy, the residuals are 
indistinguishable from noise. 
A and B have very similar colors, while G has much redder
colors (see Table~\ref{tab:0746ap}) that are similar to those of an
early-type galaxy at $z=$0.2--0.5 \citep{fukugita95}. The predicted
$I$-band magnitude of $19.3$~mag for a lens galaxy at $z_l=0.3$ 
producing the observed separation \citep{rusin03} agrees well with 
the observed magnitude of $19.6$~mag. The relative astrometry of 
the components (from the OPTIC $I$-band image) and the absolute 
photometry for the UH8k $VRI$-band observations are summarized 
in Table~\ref{tab:0746ap}.  The magnitudes are calibrated using the 
standard star PG 0231+051 \citep{landolt92}.   

We obtained 1200~sec spectra of components A and B with the Echellette 
Spectrograph and Imager \citep[ESI;][]{sutin97,sheinis02} and the MIT-LL 
2048$\times$4096 CCD detector at the Keck II telescope on 2006 March 5 
($0\farcs8$ seeing). The wavelength coverage was 3900{\,\AA} to 11,000{\,\AA} 
with a spectral dispersion of ${\sim}0\farcs25$ {\AA} pixel$^{-1}$.
The spatial sampling scale of the CCD detector is $0\farcs154$ pixel$^{-1}$ 
and we used a $1\farcs0$ wide slit aligned to observe components A and B
simultaneously. Since the spectra of the two components were well-separated
in the spectral CCD image, we were able to extract the spectra of  
the two quasars using standard IRAF\footnote{IRAF is the Image
  Reduction and Analysis Facility, a general purpose software system
  for the reduction and analysis of astronomical data. IRAF is written
  and supported by the IRAF programming group at the National Optical
  Astronomy Observatories (NOAO) in Tucson, Arizona. NOAO is operated
  by the Association of Universities for Research in Astronomy (AURA),
  Inc. under cooperative agreement with the National Science Foundation} 
tasks. We calibrated the spectra using the spectroscopic standard
G191-B2B \citep{oke90}, and the telluric lines were corrected using a 
high resolution model of the absorption bands derived from ESI 
observations of standard stars. The binned spectra of components A and B, 
whose bad pixels and bad columns were corrected by liner interpolation, 
are shown in the upper panel of Figure~\ref{fig:0746spec}, and the ratio of
the spectra is shown in the lower panel. The spectral flux ratios of
both the continuum and the \ion{Si}{4}, \ion{C}{4}, \ion{C}{3]}, and
\ion{Mg}{2} emission lines are remarkably similar, with an overall flux
ratio of ${\sim}1$.  As reported in Table~\ref{tab:0746em}, the quasar
redshifts and emission line-widths are consistent with the hypothesis 
that the spectra are identical.

%%%%%%%%%%%%%%%%%%%%%%%%%%%%%%%%%%%%%%%%%%%%%%%%
\section{Follow-Up Observations of SDSS~J1406+6126}\label{sec:1406}
%%%%%%%%%%%%%%%%%%%%%%%%%%%%%%%%%%%%%%%%%%%%%%%%

Since it was obvious that a galaxy is associated with the quasar SDSS
J1406+6126 even in the SDSS images, we acquired a short 90 sec
exposure UH8k $V$-band image on 2005 May 7 (0\farcs8 seeing) in order
to see if there are multiple stellar (quasar) images. As shown in the
upper left panel of Figure~\ref{fig:1406img}, it turned out that the
object consists of two stellar components (named components A and B,
with A being the brighter component) and an extended object (component
G) between them. We obtained a deeper and higher spatial resolution
$I$-band image with the OPTIC camera on 2006 May 3 (1\farcs0 seeing,
300 sec), which is also shown in the left column of
Figure~\ref{fig:1406img}. We again find that the residuals using a fit
consisting of only 2 PSFs are significant (the middle column of
Figure~\ref{fig:1406img}), while a model consisting of two  PSFs plus
an extended component fits the images well (the right column of
Figure~\ref{fig:1406img}).  The observed color of component G
($V-I{\sim}1.7$) is also typical of an early-type galaxy at the
spectroscopic redshift of ${z_l}=0.27$ that we discuss later in this
section \citep{fukugita95}. The relative astrometry from the OPTIC
$I$-band image and the absolute photometry from the UH8k $V$-band and
OPTIC $I$-band images for components A, B, and G are summarized in
Table~\ref{tab:1406ap}. The photometric calibration was done using the
standard stars PG 1047+003 and PG 1633+099 \citep{landolt92}. The
angular separation of components A and B is $1\farcs976\pm0\farcs013$.   

We obtained a 1200~sec spectrum of SDSS J1406+6126 on the same night
the spectra of SDSS J0746+443 were acquired, using the same
instrumental set up and observing method. Components B and G are too
close to be separated, so we extracted spectra for components A and
B+G.  The binned and bad pixel (bad column) corrected spectra of
component A and B+G are shown in Figure~\ref{fig:1406spec}. The galaxy
heavily contaminates the B+G spectrum, but component B is clearly a
quasar at the same redshift as component A and with similar emission
line widths (see Table~\ref{tab:1406em}). If we scale and subtract the
spectrum of A from the spectrum of component B+G, we find a residual
spectrum corresponding to an early-type galaxy at $z=0.27$, as shown
in Figure~\ref{fig:1406gspec}.  If we correct the B+G spectrum for the
emission from the lens galaxy, the flux ratio of the A and B spectra
shows little wavelength dependence.  

There are two interesting \ion{Mg}{2} absorption systems in the
spectra of SDSS J1406+6126. There is an absorption system at $z_{abs}=0.691$ 
(${\sim}$4730{\,\AA}) with a \ion{Mg}{2} $\lambda{2796}$ rest frame
equivalent width (REW$_{\rm MgII}$) of 1.3{\,\AA} in the spectrum of
component A that is absent (REW$_{\rm MgII}<0.3$) from the spectrum of
component B. A second system at $z_{abs}=1.562$ (${\sim}$7150{\,\AA})
has the REW$_{\rm MgII}$ of 1.3{\,\AA} in component A but only
0.3{\,\AA} in component B. These large equivalent width differences
are occurring on scales\footnote{We used the formulae quoted in
  \citet{smette92} to calculate the proper separation between the
  lensed quasar images.} of only $5.3(h/0.7)^{-1}{\rm kpc}$ (for
$z_{abs}=0.691$) and $1.1(h/0.7)^{-1}{\rm kpc}$ (for $z_{abs}=1.562$),
respectively. Such large fractional changes have been seen only for
weak absorption systems ({REW$_{\rm MgII}$}${\lesssim}0.3${\,\AA}), as
shown in Figure~10 of \citet{ellison04}. Large REW$_{\rm MgII}$
variations on kpc scales in strong intervening absorption systems,
like those we have found in the SDSS J1406+6126 spectra, could
indicate a range of size scales for such absorption systems or could
be just rare outliers.   

%%%%%%%%%%%%%%%%%%%%%%%%%%%%%%%%%%%%%%%%%%%%%%%%
%%%%%%%%%%%%%%%%%%%%%%%%%%%%%%%%%%%%%%%%%%%%%%%%
%%%%%%%%%%%%%%%%%%%%%%%%%%%%%%%%%%%%%%%%%%%%%%%%
\section{Lens Models}\label{sec:model}
%%%%%%%%%%%%%%%%%%%%%%%%%%%%%%%%%%%%%%%%%%%%%%%%
%%%%%%%%%%%%%%%%%%%%%%%%%%%%%%%%%%%%%%%%%%%%%%%%
%%%%%%%%%%%%%%%%%%%%%%%%%%%%%%%%%%%%%%%%%%%%%%%%

The follow-up data strongly supports the hypothesis that both objects
are two-image gravitational lenses. To do our final test that 
the hypothesis is reasonable, we modeled both systems using a standard
mass model consisting of a Singular Isothermal Sphere (SIS) with an
external shear.  The models have just as many parameters (the Einstein
radius $R_{\rm E}$, the shear $\gamma$ and its position angle
$\theta_\gamma$, the position of the lens galaxy, and the position and
flux of the source quasar) as the data supplies constraints (the 3
component positions and the 2 quasar fluxes), so the number of degrees
of freedom is zero. Thus we expect the mass model to be able to fit
the data perfectly, and the only check on the models is the degree to
which the parameters are physically reasonable.  We adopt the relative
positions and $I$-band flux ratios from Table~\ref{tab:0746ap} and
Table~\ref{tab:1406ap} as constraints.  

We use the {\it lensmodel} software \citep{keeton01} to fit the
models, and the parameters of the best-fitting models
($\chi^2{\sim}0$) and their 1$\sigma$ (${\Delta}{\chi^2}=1$) errors 
are summarized in Table~\ref{table:model}. The
external shears of $\gamma{\sim}0.03$ required to fit the data are
typical for lensed quasars \citep[e.g.,][]{oguri05,inada06a}.  The
predicted time delays and total magnifications are
${\Delta}t{\simeq}$3 days and $\mu_{\rm tot}{\simeq}13$ for SDSS
J0746+4403, and ${\Delta}t{\simeq}$20 days and $\mu_{\rm
  tot}{\simeq}5$ for SDSS J1406+6126. 

%%%%%%%%%%%%%%%%%%%%%%%%%%%%%%%%%%%%%%%%%%%%%%%%
%%%%%%%%%%%%%%%%%%%%%%%%%%%%%%%%%%%%%%%%%%%%%%%%
%%%%%%%%%%%%%%%%%%%%%%%%%%%%%%%%%%%%%%%%%%%%%%%%
\section{Summary}\label{sec:conc}
%%%%%%%%%%%%%%%%%%%%%%%%%%%%%%%%%%%%%%%%%%%%%%%%
%%%%%%%%%%%%%%%%%%%%%%%%%%%%%%%%%%%%%%%%%%%%%%%%
%%%%%%%%%%%%%%%%%%%%%%%%%%%%%%%%%%%%%%%%%%%%%%%%

We have discovered two new lensed quasars in the SDSS: SDSS J0746+4403
and SDSS J1406+6126.  SDSS J0746+4403 is a two-image lens
($\Delta{\theta}=1\farcs08$) formed from a $z=2.00$ quasar by a
foreground galaxy at $z_l{\sim}0.3$. The redshift of the lens galaxy
was estimated from its colors and $I$-band magnitude. SDSS J1406+6126
is a two-image lens ($\Delta{\theta}=1\farcs98$) formed from a
$z=2.13$ quasar by a spectroscopically confirmed lens galaxy at
$z_l=0.27$.  Both lenses have properties that are easily reproduced by
standard lens models with reasonable parameters.  SDSS J0746+4403 is
particularly important because it will  be included in the
well-defined statistical sample of the SQLS that we plan to use to
constrain the cosmological model. SDSS J1406+6126 is slightly fainter
than our magnitude limit for the statistical sample, but the
relatively large image separation and the bright lens galaxy should
make it a useful lens for studying the structure of early-type
galaxies. A particularly interesting feature of SDSS J1406+6126 is
that it has two strong \ion{Mg}{2} absorption line systems that show
dramatic changes on the few kpc scales of the path separations, which
could provide an useful insight into the size distribution of
absorption systems. 
Spectroscopy to measure the redshifts and velocity dispersions of the
lens galaxies, high-resolution imaging to determine the structures of
the lens galaxies, and measurements of the time delays are the next steps
towards using these lenses to study the structure and evolution of
galaxies.

\acknowledgments

Use of the UH 2.2-m telescope for the observations is supported by
NAOJ. Some of the data presented herein were obtained at the W.M. Keck
Observatory, which is operated as a scientific partnership among the
California Institute of Technology, the University of California and
the National Aeronautics and Space Administration. The Observatory was
made possible by the generous financial support of the W.M. Keck
Foundation. This work is also based in part on observations obtained
with the MDM 2.4m Hiltner, which is owned and operated by a consortium
consisting of Columbia University, Dartmouth College, the University
of Michigan, the Ohio State University and Ohio University.   We would
like to thank D. Depoy, J. Eastman, S. Frank, J. Marshall of OSU and
J. Halpern of Columbia University for operating the MDM queue
observing and monitoring program. 

N.~I. is supported by JSPS through JSPS Research Fellowship for Young
Scientists. This work was supported in part by the Department of
Energy contract DE-AC02-76SF00515. A portion of this work was also
performed under the auspices of the U.S. Department of Energy,
National Nuclear Security Administration by the University of
California, Lawrence Livermore National Laboratory under contract
No. W-7405-Eng-48.  I.~K. acknowledges the support from Ministry of
Education, Culture, Sports, Science, and Technology, Grant-in-Aid for
Encouragement of Young Scientists (No. 17740139).

Funding for the SDSS and SDSS-II has been provided by the Alfred
P. Sloan Foundation, the Participating Institutions, the National
Science Foundation, the U.S. Department of Energy, the National
Aeronautics and Space Administration, the Japanese Monbukagakusho, the
Max Planck Society, and the Higher Education Funding Council for
England. The SDSS Web Site is http://www.sdss.org/. 

The SDSS is managed by the Astrophysical Research Consortium for the
Participating Institutions. The Participating Institutions are the
American Museum of Natural History, Astrophysical Institute Potsdam,
University of Basel, Cambridge University, Case Western Reserve
University, University of Chicago, Drexel University, Fermilab, the
Institute for Advanced Study, the Japan Participation Group, Johns
Hopkins  University, the Joint Institute for Nuclear Astrophysics, the
Kavli Institute for Particle Astrophysics and Cosmology, the Korean
Scientist Group, the Chinese Academy of Sciences (LAMOST), Los Alamos
National Laboratory, the Max-Planck-Institute for Astronomy (MPIA),
the Max-Planck-Institute for Astrophysics (MPA), New Mexico State
University, Ohio State University, University of Pittsburgh,
University of Portsmouth, Princeton University, the United States
Naval Observatory, and the University of Washington.

{\it Facilities:} \facility{SDSS 2.5-m}, \facility{Keck II (ESI)}, 
\facility{UH88 (UH8k, OPTIC)}, \facility{MDM 2.4-m (RETROCAM)}.

\clearpage

%%%%%%%%%%%%%%%%%%%%%%%%%%%%%%%%%%%%%%%%%%%%%%%%%%%%%%%%%%%%%%%%%%%%%%%
%%%%%%%%%%%%%%%%%%%%%%%%%%%%%%%%%%%%%%%%%%%%%%%%%%%%%%%%%%%%%%%%%%%%%%%

\clearpage

%%%%%%%%%%%%%%%%%%%%%%%%%%%%%%%%%%%%%%%%%%%%%%%%%%%%%%%%%%%%%%%%%%%%%%%
\begin{figure}
\epsscale{.8}
\plotone{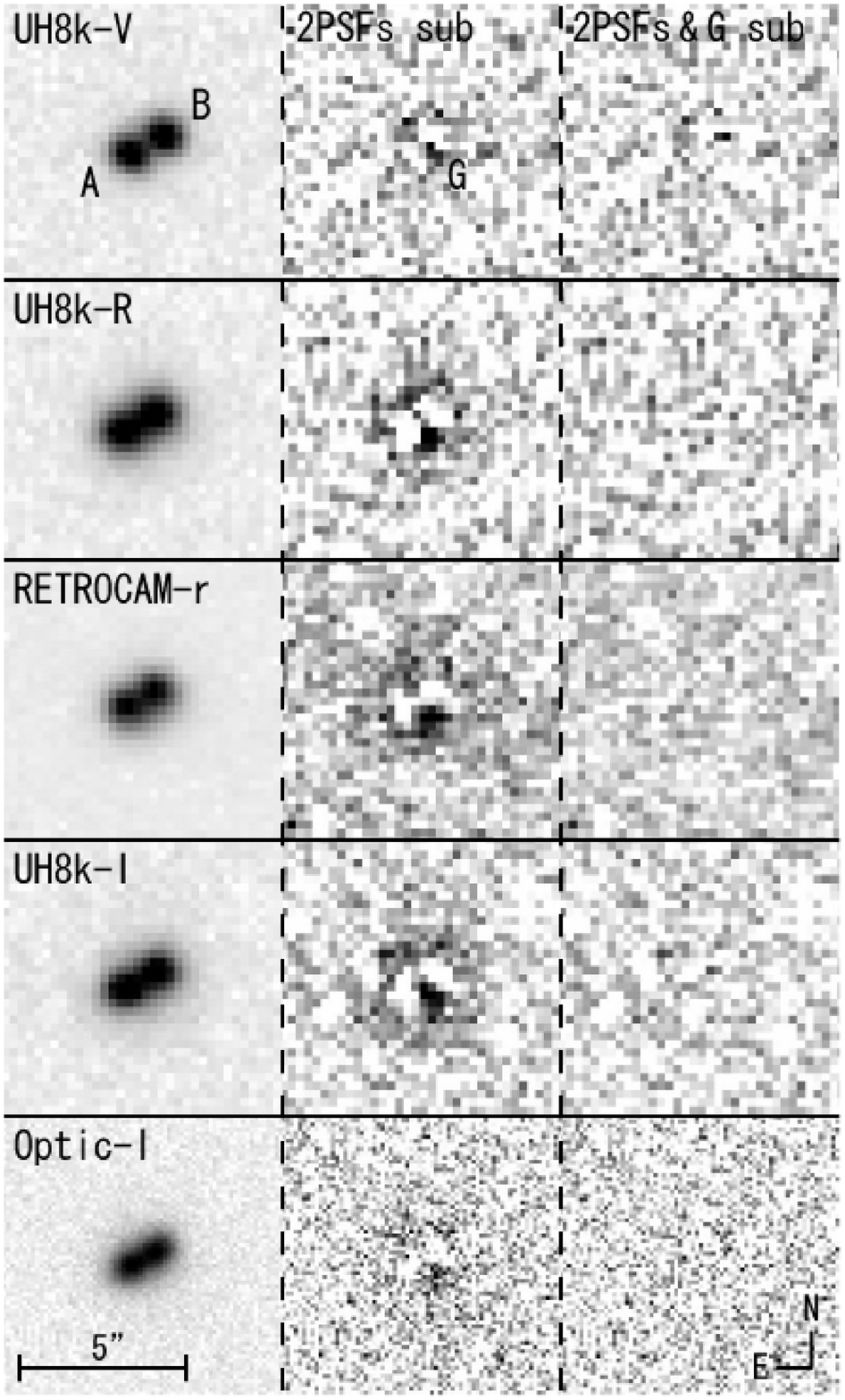}
\end{figure}
%%%%%%%%%%%%%%%%%%%%%%%%%%%%%%%%%%%%%%%%%%%%%%%%%%%%%%%%%%%%%%%%%%%%%%%

\clearpage

%%%%%%%%%%%%%%%%%%%%%%%%%%%%%%%%%%%%%%%%%%%%%%%%%%%%%%%%%%%%%%%%%%%%%%%
\begin{figure}
\caption{The UH88 8k mosaic CCD $VRI$-band images, the MDM 2.4-m
 RETROCAM $r$-band image, and  the UH88 OPTIC $I$-band image of SDSS
 J0746+4403. The image scales are 0\farcs235 ${\rm pixel^{-1}}$ for the
 UH8k, 0\farcs259 ${\rm pixel^{-1}}$ for the RETROCAM, and 0\farcs137
 ${\rm pixel^{-1}}$ for the OPTIC. The left panels show the original
 images. In the middle panels we show the residuals after fitting the 
 images using {\it only} 2 PSFs, and the right panels show the residuals
 after fitting the images using 2 PSFs
 plus 1 galaxy component. Residual fluxes, which originate from the lens
 galaxy (component G), can be seen in all the middle panels, and no 
 residuals can be seen in any of the right panels. The colors of 
 components A and B are quite similar (see Table~\ref{tab:0746ap}), 
 and the colors and magnitude of component G are consistent with 
 those of an early-type galaxy at $z{\sim}0.3$.  
\label{fig:0746img}}
\end{figure}
%%%%%%%%%%%%%%%%%%%%%%%%%%%%%%%%%%%%%%%%%%%%%%%%%%%%%%%%%%%%%%%%%%%%%%%

\clearpage

%%%%%%%%%%%%%%%%%%%%%%%%%%%%%%%%%%%%%%%%%%%%%%%%%%%%%%%%%%%%%%%%%%%%%%%
\begin{figure}
\epsscale{.85}
\plotone{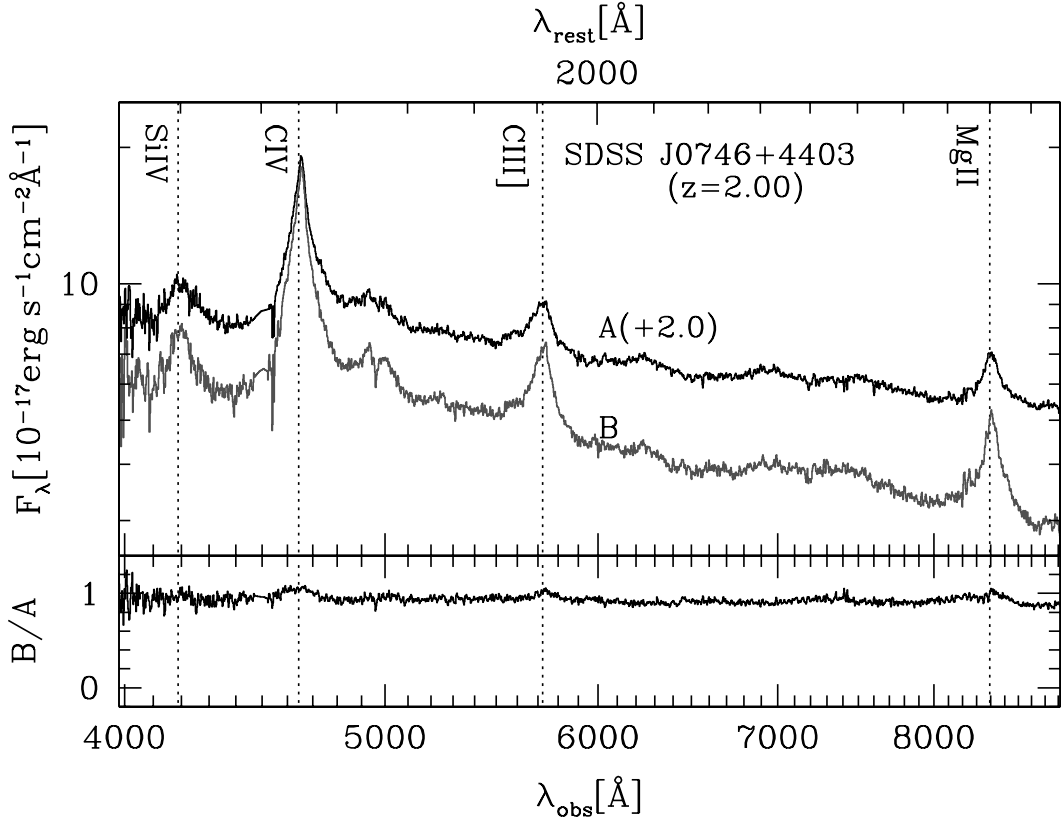}
\caption{Binned and bad pixel (bad column) corrected 
  spectra of components A ({\sl black} solid line, 
  shifted upwards by a constant of $2.0{\times}10^{-17}$) 
  and B ({\sl gray} solid line) of SDSS 
  J0746+4403, taken with the ESI at the Keck II telescope. 
  The original spectra has the reciprocal dispersion of 
  $\sim$11.4 km ${\rm s}^{-1}$ ${\rm pixel}^{-1}$. 
  The vertical dotted lines indicate the positions of the 
  quasar \ion{Si}{4}, \ion{C}{4}, \ion{C}{3]}, and \ion{Mg}{2} emission 
  lines redshifted to $z=2.00$. In addition to the emission lines, 
  there is a common \ion{C}{4} absorption system at $\sim$4540{\,\AA} 
  ($z_{abs}=1.93$)
  in the spectra of components A and B, and there is a weak 
  (REW$_{\rm MgII}{\lesssim}0.3$) \ion{Mg}{2} absorption system 
  at $\sim$7430{\,\AA} ($z_{abs}=1.65$) in the spectrum of component A. 
  The spectral flux ratio of components 
  A and B is nearly constant, as shown in the lower panel.
\label{fig:0746spec}}
\end{figure}
%%%%%%%%%%%%%%%%%%%%%%%%%%%%%%%%%%%%%%%%%%%%%%%%%%%%%%%%%%%%%%%%%%%%%%%

\clearpage

%%%%%%%%%%%%%%%%%%%%%%%%%%%%%%%%%%%%%%%%%%%%%%%%%%%%%%%%%%%%%%%%%%%%%%%
\begin{figure}
\epsscale{.8}
\plotone{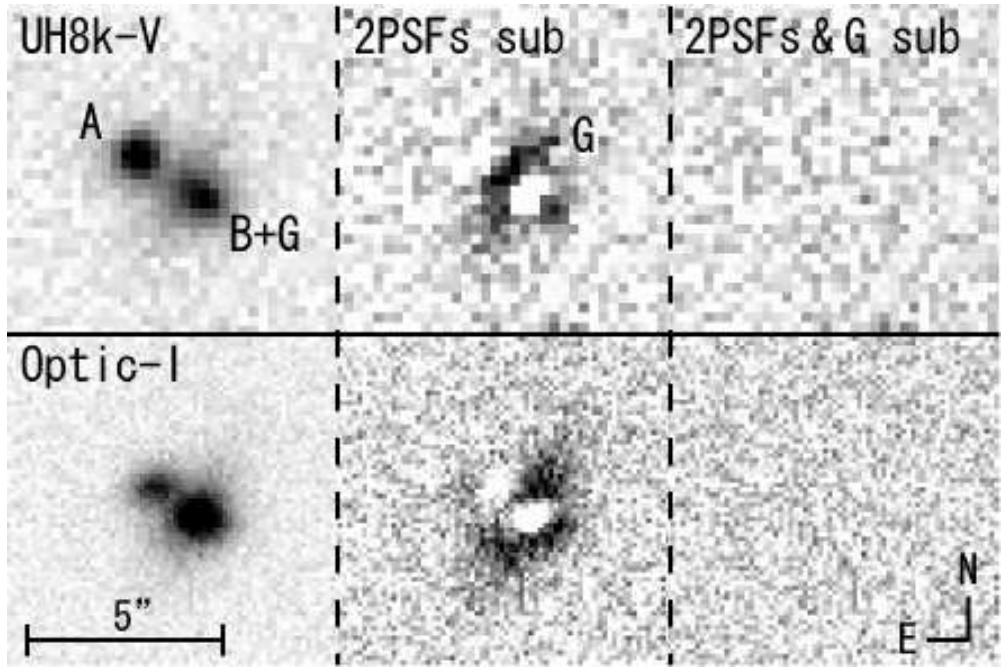}
\caption{The UH8k $V$-band image and the OPTIC $I$-band image of SDSS J1406+6126. 
  The format of the figure is the same as in Figure~\ref{fig:0746img}. 
  An extended object can be seen between the two stellar components even 
  in the original images, and its presence is obvious in the middle panels
  showing the residuals from a fit consisting only of 2 PSFs, while the  
  model consisting of two PSFs plus an extended galaxy fits the data
  very well (right panels).  The colors of components A and B are
  quite similar (Table~\ref{tab:1406ap}), and the color of component G is
  consistent with that of an early-type galaxy at $z_l=0.2$, which is in
  good agreement with its spectroscopic redshift of $z_l=0.27$.
\label{fig:1406img}}
\end{figure}
%%%%%%%%%%%%%%%%%%%%%%%%%%%%%%%%%%%%%%%%%%%%%%%%%%%%%%%%%%%%%%%%%%%%%%%

\clearpage

%%%%%%%%%%%%%%%%%%%%%%%%%%%%%%%%%%%%%%%%%%%%%%%%%%%%%%%%%%%%%%%%%%%%%%%
\begin{figure}
\epsscale{.85}
\plotone{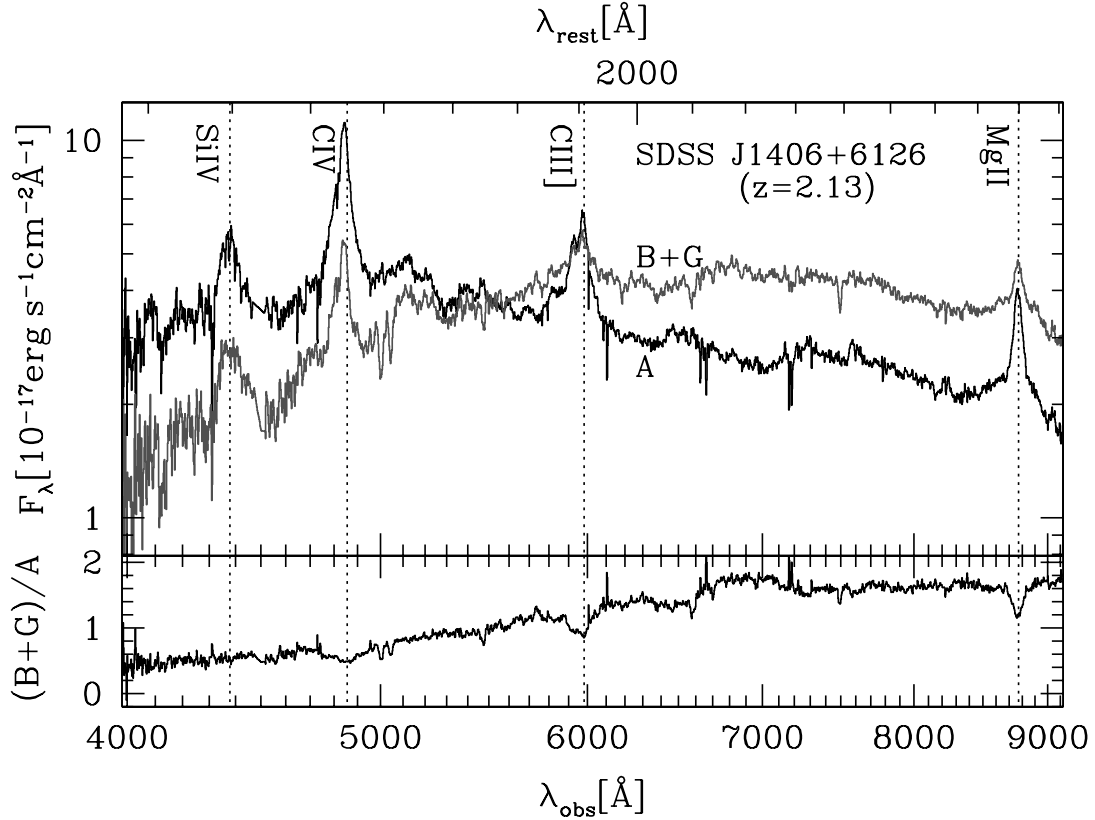}
\caption{Binned and bad pixel (bad column) corrected  spectra of
 components A ({\sl black} solid line) and the combination of B+G
 ({\sl gray} solid line) of SDSS J1406+6126, taken with the ESI at the
 Keck II telescope. The original spectra has the reciprocal dispersion
 of  $\sim$11.4 km ${\rm s}^{-1}$ ${\rm pixel}^{-1}$.  The vertical
 dotted lines indicate the positions of the quasar \ion{Si}{4},
 \ion{C}{4}, \ion{C}{3]}, and \ion{Mg}{2} emission lines redshifted to
 $z=2.13$.  In addition to the interesting \ion{Mg}{2} absorption systems 
 (see \S \ref{sec:1406}),  there are two common \ion{C}{4} absorption
 systems at $\sim$4320{\,\AA} ($z_{abs}=1.78$) and $\sim$4130{\,\AA}
 ($z_{abs}=1.66$) in the spectra of components A and B, and one of the
 interesting \ion{Mg}{2} absorption systems (at $z_{abs}=1.56$) has
 \ion{Fe}{2} absorption lines (at $\sim$6660{\,\AA},
 $\sim$6630{\,\AA}, $\sim$6100{\,\AA}, $\sim$6080{\,\AA}, and
 $\sim$6000{\,\AA}) and a \ion{Si}{2} absorption line (at
 $\sim$4650{\,\AA}). The spectral flux ratio between components A and
 B+G is shown in the lower panel.  It is not constant because of the
 flux of component G. However, both the spectra clearly have quasar
 emission lines at the same redshift and with the similar widths (see
 Table~\ref{tab:1406em}). 
\label{fig:1406spec}}
\end{figure}
%%%%%%%%%%%%%%%%%%%%%%%%%%%%%%%%%%%%%%%%%%%%%%%%%%%%%%%%%%%%%%%%%%%%%%%

\clearpage

%%%%%%%%%%%%%%%%%%%%%%%%%%%%%%%%%%%%%%%%%%%%%%%%%%%%%%%%%%%%%%%%%%%%%%%
\begin{figure}
\epsscale{.85}
\plotone{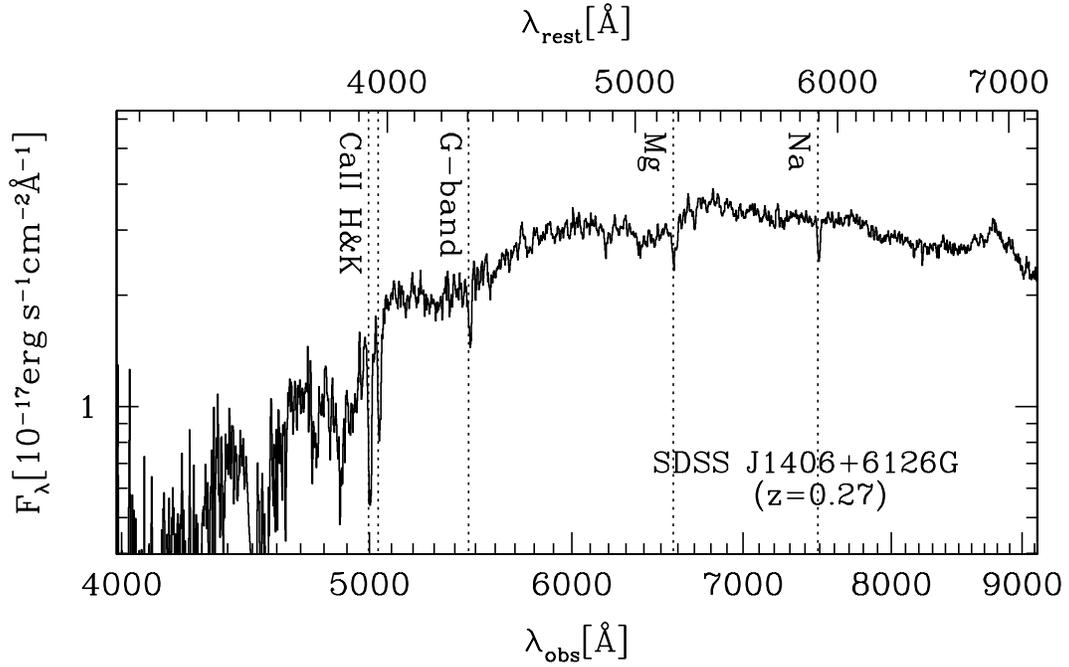}
\caption{The spectrum of component G of SDSS J1406+6126 found by
 subtracting $0.4$ times the spectrum of component A from that of
 components B+G.  The spectrum is that of an early-type galaxy at
 $z=0.27$. The vertical dotted lines indicate the positions of the
 lens galaxy absorption lines (\ion{Ca}{2} H \& K, G-band, Mg, and Na)
 redshifted to $z=0.27$. 
\label{fig:1406gspec}}
\end{figure}
%%%%%%%%%%%%%%%%%%%%%%%%%%%%%%%%%%%%%%%%%%%%%%%%%%%%%%%%%%%%%%%%%%%%%%%

\clearpage

%%%%%%%%%%%%%%%%%%%%%%%%%%%%%%%%%%%%%%%%%%%%%%%%%%%%%%%%%%%%%%%%%%%%%%%
\begin{deluxetable}{ccccccc}
\tabletypesize{\normalsize}
\rotate
\tablecaption{SDSS PHOTOMETRY AND REDSHIFTS OF LENSES\label{tab:sdss}}
\tablewidth{0pt}
\tablehead{ \colhead{Object} & \colhead{$u$\tablenotemark{a}} &
\colhead{$g$\tablenotemark{a}} & \colhead{$r$\tablenotemark{a}} & 
\colhead{$i$\tablenotemark{a}} & \colhead{$z$\tablenotemark{a}} & 
\colhead{Redshift\tablenotemark{b}} }
\startdata
SDSS J0746+4403 & 18.83{$\pm$}0.03 & 18.74{$\pm$}0.01 & 18.70{$\pm$}0.02 & 18.41{$\pm$}0.02 & 18.20{$\pm$}0.06 & 1.998{$\pm$}0.002 \\
SDSS J1406+6126 & 19.70{$\pm$}0.07 & 19.37{$\pm$}0.06 & 19.21{$\pm$}0.14 & 19.11{$\pm$}0.18 & 18.78{$\pm$}0.22 & 2.134{$\pm$}0.002 \\
\enddata
\tablenotetext{a}{total magnitudes inside a ${\sim}2\farcs0$ aperture radius 
and without Galactic extinction corrections from the SDSS data.} 
\tablenotetext{b}{quasar redshifts from the SDSS spectra.} 
\end{deluxetable}
%%%%%%%%%%%%%%%%%%%%%%%%%%%%%%%%%%%%%%%%%%%%%%%%%%%%%%%%%%%%%%%%%%%%%%%

\clearpage

%%%%%%%%%%%%%%%%%%%%%%%%%%%%%%%%%%%%%%%%%%%%%%%%%%%%%%%%%%%%%%%%%%%%%%%
\begin{deluxetable}{crrccc}
\tabletypesize{\normalsize}
\tablecaption{ASTROMETRY AND PHOTOMETRY OF SDSS~J0746+4403\label{tab:0746ap}}
\tablewidth{0pt}
\tablehead{\colhead{Component} & \colhead{{$\Delta$}{\rm X} (arcsec)\tablenotemark{a}} &
 \colhead{{$\Delta$}{\rm Y} (arcsec)\tablenotemark{a}} & 
 \colhead{$V$\tablenotemark{b}} & \colhead{$R$\tablenotemark{b}} & \colhead{$I$\tablenotemark{b}} }
\startdata
A &  0.000{$\pm$}0.003 & 0.000{$\pm$}0.003     & 19.87{$\pm$}0.01 & 19.53{$\pm$}0.01 & 19.08{$\pm$}0.01 \\
B &  $-$0.933{$\pm$}0.004 & 0.541{$\pm$}0.004  & 19.97{$\pm$}0.05 & 19.55{$\pm$}0.02 & 19.11{$\pm$}0.01 \\
G &  $-$0.597{$\pm$}0.041 & 0.201{$\pm$}0.041  & 21.57{$\pm$}1.00 & 20.65{$\pm$}0.05 & 19.62{$\pm$}0.19 \\
\enddata
\tablenotetext{a}{Measured in the OPTIC $I$-band image using
 GALFIT. The positive directions of X and Y are defined by east and
 north, respectively. The celestial coordinates of component A are
 R.A.$=116\fdg72113$ and Decl.$=+44\fdg06416$ (J2000).} 
\tablenotetext{b}{Measured in the UH8k $VRI$-band images using
 GALFIT. The errors do not include the photometric uncertainty of the
 standard star.} 
\end{deluxetable}
%%%%%%%%%%%%%%%%%%%%%%%%%%%%%%%%%%%%%%%%%%%%%%%%%%%%%%%%%%%%%%%%%%%%%%%

\clearpage

%%%%%%%%%%%%%%%%%%%%%%%%%%%%%%%%%%%%%%%%%%%%%%%%%%%%%%%%%%%%%%%%%%%%%%%
\begin{deluxetable}{lccccccc} 
\rotate
\tablecolumns{8} 
\tablewidth{0pc} 
\tablecaption{EMISSION LINES OF SDSS~J0746+4403 SPECTRA\label{tab:0746em}} 
\tablehead{ 
\colhead{} & \multicolumn{3}{c}{Component A} & \colhead{} & 
\multicolumn{3}{c}{Component B} \\ 
\cline{2-4} \cline{6-8} \\ 
\colhead{Line({\,\AA})} & \colhead{${\lambda}_{obs}$({\,\AA})} & \colhead{FWHM({\,\AA})} & \colhead{Redshift} & \colhead{} &
\colhead{${\lambda}_{obs}$({\,\AA})} & \colhead{FWHM({\,\AA})} & \colhead{Redshift} }
\startdata 
\ion{Si}{4} (1396.76) & 4196.90 & 107.5 & 2.0047$\pm$0.0015 & & 4197.36  & 94.3  & 2.0051$\pm$0.0008 \\
\ion{C}{4} (1549.06)  & 4654.18 & 95.8  & 2.0045$\pm$0.0004 & & 4654.32  & 96.7  & 2.0046$\pm$0.0002 \\
\ion{C}{3]} (1908.73) & 5725.77 & 119.2 & 1.9998$\pm$0.0009 & & 5726.97  & 113.3 & 2.0004$\pm$0.0006 \\
\ion{Mg}{2} (2798.75) & 8402.93 & 155.7 & 2.0024$\pm$0.0005 & & 8403.06  & 166.5 & 2.0024$\pm$0.0007 \\
\enddata 
\end{deluxetable} 
%%%%%%%%%%%%%%%%%%%%%%%%%%%%%%%%%%%%%%%%%%%%%%%%%%%%%%%%%%%%%%%%%%%%%%%

\clearpage

%%%%%%%%%%%%%%%%%%%%%%%%%%%%%%%%%%%%%%%%%%%%%%%%%%%%%%%%%%%%%%%%%%%%%%%
\begin{deluxetable}{crrcc}
\tabletypesize{\normalsize}
\tablecaption{ASTROMETRY AND PHOTOMETRY OF SDSS~J1406+6126\label{tab:1406ap}}
\tablewidth{0pt}
\tablehead{\colhead{Object} & \colhead{{$\Delta$}{\rm X} (arcsec)\tablenotemark{a}} &
 \colhead{{$\Delta$}{\rm Y} (arcsec)\tablenotemark{a}} & 
 \colhead{$V$\tablenotemark{b}} & \colhead{$I$\tablenotemark{b}} }
\startdata
A &  0.000{$\pm$}0.004 & 0.000{$\pm$}0.004        & 19.99{$\pm$}0.01 & 19.38{$\pm$}0.01 \\
B &  $-$1.639{$\pm$}0.007 & $-$1.103{$\pm$}0.007  & 20.60{$\pm$}0.03 & 19.97{$\pm$}0.04 \\
G &  $-$1.143{$\pm$}0.006 & $-$0.727{$\pm$}0.006  & 19.86{$\pm$}0.07 & 18.12{$\pm$}0.02 \\
\enddata
\tablenotetext{a}{Measured in the OPTIC $I$-band image using
 GALFIT. The positive directions of X and  Y are defined by east and
 north, respectively. The celestial coordinates of component A are
 R.A.$=211\fdg60344$ and Decl.$=+61\fdg44474$ (J2000).} 
\tablenotetext{b}{Measured in the UH8k $V$-band image and OPTIC
 $I$-band image using GALFIT. The errors do not include the photometric
 uncertainty of the standard star.} 
\end{deluxetable}
%%%%%%%%%%%%%%%%%%%%%%%%%%%%%%%%%%%%%%%%%%%%%%%%%%%%%%%%%%%%%%%%%%%%%%%

\clearpage

%%%%%%%%%%%%%%%%%%%%%%%%%%%%%%%%%%%%%%%%%%%%%%%%%%%%%%%%%%%%%%%%%%%%%%%
\begin{deluxetable}{lccccccc} 
\rotate
\tablecolumns{8} 
\tablewidth{0pc} 
\tablecaption{EMISSION LINES OF SDSS~J1406+6126 SPECTRA\label{tab:1406em}} 
\tablehead{ 
\colhead{} & \multicolumn{3}{c}{Component A} & \colhead{} & 
\multicolumn{3}{c}{Component B} \\ 
\cline{2-4} \cline{6-8} \\ 
\colhead{Line({\,\AA})} & \colhead{${\lambda}_{obs}$({\,\AA})} & \colhead{FWHM({\,\AA})} & \colhead{Redshift} & \colhead{} &
\colhead{${\lambda}_{obs}$({\,\AA})} & \colhead{FWHM({\,\AA})} & \colhead{Redshift} }
\startdata 
\ion{Si}{4}(1396.76) & 4376.51 & 90.6  & 2.1333$\pm$0.0009 & & 4377.14  & 95.7  & 2.1338$\pm$0.0009 \\
\ion{C}{4}(1549.06)  & 4842.11 & 73.9  & 2.1258$\pm$0.0004 & & 4841.59  & 63.2  & 2.1255$\pm$0.0005 \\
\ion{C}{3]}(1908.73) & 5975.09 & 73.6  & 2.1304$\pm$0.0005 & & 5974.41  & 69.5  & 2.1300$\pm$0.0012 \\
\ion{Mg}{2}(2798.75) & 8766.79 & 83.4  & 2.1324$\pm$0.0004 & & 8767.50  & 97.4  & 2.1326$\pm$0.0008 \\
\enddata 
\end{deluxetable} 
%%%%%%%%%%%%%%%%%%%%%%%%%%%%%%%%%%%%%%%%%%%%%%%%%%%%%%%%%%%%%%%%%%%%%%%

\clearpage

%%%%%%%%%%%%%%%%%%%%%%%%%%%%%%%%%%%%%%%%%%%%%%%%%%%%%%%%%%%%%%%%%%%%%%%
\begin{deluxetable}{cccccc}
\tablewidth{0pt}
\tablecaption{RESULTS OF LENS MODELING\label{table:model}}
\tablehead{\colhead{Object} & \colhead{$R_{\rm E}$(${}''$)} &
 \colhead{$\gamma$} &
 \colhead{$\theta_\gamma{({}^{\circ})}$\tablenotemark{a}} & \colhead{$\Delta
 t$[$h^{-1}$day]} & \colhead{{$\mu_{\rm tot}$} \tablenotemark{c}}} 
\startdata
SDSS J0746+4403  & $0.55\pm$0.01 & $0.034\pm$0.025 & $-4.7\pm$5.7 & $2.4$\tablenotemark{b} & 13.2 \\
SDSS J1406+6126  & $0.97\pm$0.01 & $0.027\pm$0.008 & $-17.2\pm$5.5 & $18.8$ & 5.2 \\
\enddata
\tablenotetext{a}{Each position angle is measured East of North.}
\tablenotetext{b}{The lens galaxy redshift of SDSS J0746+4403 is assumed to be $z=0.3$.}
\tablenotetext{c}{Predicted total magnification.}
\end{deluxetable}
%%%%%%%%%%%%%%%%%%%%%%%%%%%%%%%%%%%%%%%%%%%%%%%%%%%%%%%%%%%%%%%%%%%%%%%

\end{document}